\documentclass[conference]{IEEEtran}
\IEEEoverridecommandlockouts
\usepackage{color}
\usepackage[pdftex]{graphicx}
\usepackage{amsmath}
\usepackage{amssymb}
\usepackage{hyperref}

\begin{document}

\title{Application of Cybernetics and Control \\ Theory for a New Paradigm in Cybersecurity\IEEEauthorrefmark{1}}

\author{\IEEEauthorblockN{Michael D. Adams}
	\IEEEauthorblockA{L-3 Communications, CSW \\ 
	Salt Lake City, Utah 84116 \\
	michael.d.adams@L-3com.com} 
	\thanks{\IEEEauthorrefmark{1} ``The information in this paper consists of general capabilities data from public sources and is not subject to any International Traffic in Arms Regulations (ITAR) or Export Administration Regulations (EAR) controls.''}
	\and
\IEEEauthorblockN{Seth D. Hitefield}
	\IEEEauthorblockA{Hume Center, Virginia Tech\\
	Blacksburg, Virginia 24061\\
	seth.hitefield@vt.edu} \and
\IEEEauthorblockN{Bruce Hoy}
	\IEEEauthorblockA{L-3 Communications, NSS \\
	Reston, Virginia 20190 \\
	bruce.hoy@L-3Com.com} \and
\IEEEauthorblockN{Michael C. Fowler}
	\IEEEauthorblockA{Hume Center, Virginia Tech \\
	Blacksburg, Virginia 24061\\
	mifowler@vt.edu} \and
\IEEEauthorblockN{T. Charles Clancy}
	\IEEEauthorblockA{Hume Center, Virginia Tech \\
	Arlington, Virginia 22203\\
	tcc@vt.edu}
}


%

\maketitle

\begin{abstract}
A significant limitation of current cyber security research and techniques is its reactive and applied nature. This leads to a continuous `cyber cycle' of attackers scanning networks, developing exploits and attacking systems, with defenders detecting attacks, analyzing exploits and patching systems. This reactive nature leaves sensitive systems highly vulnerable to attack due to un-patched systems and undetected exploits. Some current research attempts to address this major limitation by introducing systems that implement moving target defense. However, these ideas are typically based on the intuition that a moving target defense will make it much harder for attackers to find and scan vulnerable systems, and not on theoretical mathematical foundations. The continuing lack of fundamental science and principles for developing more secure systems has drawn increased interest into establishing a `science of cyber security'. This paper introduces the concept of using cybernetics, an interdisciplinary approach of control theory, systems theory, information theory and game theory applied to regulatory systems, as a foundational approach for developing cyber security principles. It explores potential applications of cybernetics to cyber security from a defensive perspective, while suggesting the potential use for offensive applications. Additionally, this paper introduces the fundamental principles for building non-stationary systems, which is a more general solution than moving target defenses. Lastly, the paper discusses related works concerning the limitations of moving target defense and one implementation based on non-stationary principles.
\end{abstract}


\IEEEpeerreviewmaketitle


\section{Introduction}
	\label{sec:introduction}
	With computer networks now providing the backbone for critical systems in financial centers, heath care, governments, militaries, public infrastructures and other industries worldwide, cyber security has emerged as a significant research area for academia, industry and government. Developing more secure systems is paramount as computers are further integrated into everyday life and store increasing amounts of sensitive data. However, a significant limitation of current security research is that the many solutions, both past and present, are a purely reactive response to known vulnerabilities and attacks. This leads to implementations that mitigate these vulnerabilities, using techniques such as patching, encryption and firewalls, but still leave sensitive systems exposed to undiscovered attacks. In addition, innovative solutions used to solve past security issues can quickly become obsolete as technology improves and matures.

\begin{figure}[]
  \centering
  \includegraphics[width=0.48\textwidth]{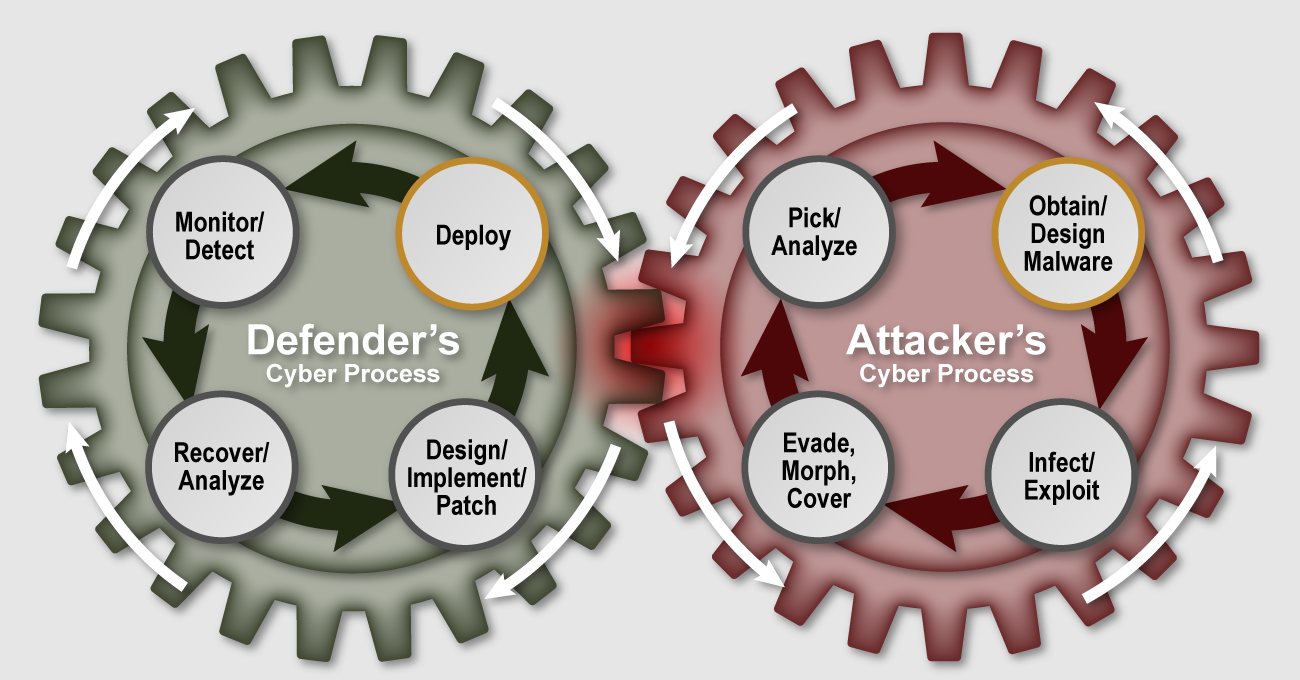}
  \caption{Cyber Cycle}
  \label{fig:cyber_cycle}
\end{figure}

Relying on a reactive methodology for improving cyber security has resulted in what some have coined as an `arms race' \cite{armstrong2009complexity,mcmorrow2010cybersec,goodman2007research,daly2002deming} or also a `cyber cycle' as shown in Figure \ref{fig:cyber_cycle}. Essentially, the current state of cyber security can be viewed as a continuous game between network attackers and defenders. On one side, a defender designs and deploys a system on its network to provide a service to its users. Concurrently, the attackers are scanning that network, analyzing any available systems in order to determine their properties and any other types of communication channels or vulnerabilities that might exist. Once the attacker has discovered a vulnerable target (typically a result of poor security design or coding flaw), their next step is to obtain malware that exploits those vulnerabilities. At this point, the defender is monitoring the network (using tools such as intrusion detection and firewall systems) attempting to detect any successful attacks on the network. If the attacker is successful in exploiting discovered vulnerabilities, they can use the unauthorized access to their benefit, which may include data exfiltration, denial of service for other users, performance degradation, system destruction, etc. If and when the defender detects the attack on the system, they must first recover the system to a stable state and then analyze the affected systems and malware to determine what vulnerabilities were exploited. After this, they must design a new patch or security strategy to protect the system from further exploit, which results in the cycle repeating with attackers rescanning and reanalyzing the systems, searching for new vulnerabilities \cite{mcmorrow2010cybersec,daly2002deming}.

Responding to attacks in this manner is not sufficient; security research must have a fundamental change in its focus to a balance of theoretical and applied research designed to disrupt the ``cyber cycle'' \cite{mcmorrow2010cybersec}. Past approaches tended to focus on application and protocol-specific vulnerabilities and were typically applied solutions that did not consider fundamental issues. Good examples of such solutions are HTTPS, DNSSEC, and two-factor authentication, which solve simple problems such as authentication integrity and insecure communication, but do not solve major fundamental problems like insecure web servers, DNS cache poisoning, and breakable passwords \cite{honan2012}. In each case, the solution provided greater security for a specific vulnerability, but the main protocols were essentially unchanged and did not change the fact the systems remain highly vulnerable. Applied solutions alone cannot solve the larger fundamental issues in cyber security, and new research must be built on fundamental mathematical principles.

Some engineers and scientists in the areas of network and computer security have recognized the need for a change in security research. However, there is still a division between applied and fundamental research that needs to be addressed. For example, one popular approach with many researchers, further discussed in section \ref{sec:related_works}, is the principle of moving targets applied to network addressing, configuration, and web-server security and reliability \cite{huang2009automating,dunlop2011mt6d}. But, these projects still lack fundamental principles needed to create more secure systems. Moving target defense research is mostly based on the correct intuition that moving targets are more difficult to hit. On the other hand, there have been several groups within the government, including the Departments of Defense, Energy, and Homeland Security, that have recognized the need for a more fundamental scientific approach. In 2005, the President's Information Technology Advisory Committee (PITAC) released their report \cite{benioff2005cyber}, which recognized that add-on security and continuous software patches are inadequate to protect the significant network infrastructure in the government and acknowledged the need for longer-term fundamental research that would provide a basis for future security research. The 2010 report \cite{mcmorrow2010cybersec} from the JASON group in MITRE stated the need for a combination of applied and theoretical research and also discussed how applications from other sciences could help build up a `science of cyber-security'.   A chapter written by Dr. Alexander Kott in the Springer series \emph{Advances in Information Security} \cite{kott2014towards}, went as far as developing semi-formal equations and definitions with the understanding that malware is a significant problem of cyber-security. Similar papers have developed fundamental definitions and equations for describing the attack surfaces of a system \cite{jajodia2011moving,jajodia2012moving}. This type of research work has identified that deficiencies exist, but has been unable to demonstrate the mathematical foundation required to quantitatively measure the effectiveness of a cyber security solution. The application of cybernetics and control theory will attempt to fill this void and provide a new framework established from proven mathematical constructs that will serve as the basis to develop a new class of cyber security solutions.

The purpose of this paper is to introduce the science of cybernetics as a fundamental base for security research. Even though cybernetics covers multiple disciplines including control theory, systems theory, information theory and game theory, this paper will specifically focus on the control theory aspects of cybernetics. Using the concepts of feedback and system control, better theoretical approaches and mathematical principles can be developed and deployed to improve applied security solutions. As mentioned in the JASON report, security research should be a combination of both applied and theoretical approaches. Using cybernetics as a starting point for security research develops fundamental models that can be quickly applied to current security solutions. In some cases, those models can be used to validate and improve moving target systems currently being developed. In other cases, using a cybernetic-based approach provides researchers with a different perspective from which to analyze security issues and develop models and solutions that help break the continuous cyber cycle.

The remainder of this paper is organized as follows. Section \ref{sec:why_cybernetics} quickly introduces the motivation for applying cybernetic principles to cyber-security. Section \ref{sec:cybernetics_overview} provides a basic introduction of several cybernetic concepts which are applied to a cybernetic model of the cybercycle in Section \ref{sec:cybernetic_applications}. In addition, this section further explores cybernetic applications to cyber-security and provides several insights into developing a science of cyber-security. Finally, Section \ref{sec:related_works} discusses some existing implementions 


\section{Why Cybernetics?}
	\label{sec:why_cybernetics}
	Claude Shannon, the recognized father of information theory, is quoted as having said in the 1940s to Norbert Wiener, the father of cybernetics, ``Use the word `cybernetics' Norbert, because nobody knows what it means. This will always put you at an advantage in arguments \cite{cyberneticdef}.

Indeed, as we researched our ideas to discover whether there were others that had written on the connection between cybernetics and cyber security, we discovered no academic literature specifically applying cybernetic concepts to cyber security. Because of this perceived lack of domain knowledge and the fact that the pioneers of the subject themselves admit it is a somewhat obscure topic, we will first attempt to explain a few tenets of the subject as an introduction to our thesis. In doing so, however, we will ignore more recent developments in cybernetics such as second and third-order cybernetics and concentrate more on the root of the cybernetic development tree reaching back to the 1940s and 1950s. In doing so, we recognize we may be ignoring areas directly applicable to our thesis. It is our hope, however, to spawn creativity in individuals with advanced expertise in cybernetics to contribute further to our efforts in establishing cybernetics as a foundation for the `science of cyber security'.

So, what is cybernetics? Cybernetics is the regulation of systems \cite{gwu}. It was born in the pre-digital computer era when a handful of people saw the need for the science of regulating systems. This is not unlike our need to develop a science of cyber security today. As they began developing the principles and tenets of the discipline, they discovered that these logical and mathematical principles, not being tied to any specific physical embodiment, applied equally well from the molecule to the galaxy. The term cybernetics, derived from the Greek word `$\kappa\upsilon\beta\epsilon\nu\eta\tau\eta\varsigma$' meaning `steersman, governor, pilot, or rudder,' was used by Norbert Wiener for his 1948 introductory text on the subject \cite{wiener1948cybernetics}. In 1957, subsequent to Wiener's work, William Ashby added to the domain by publishing his often-cited book, ``An Introduction to Cybernetics \cite{ashby1955introduction}.'' We will refer a great deal in section \ref{sec:cybernetics_overview} of this paper to this well-written text. Therefore, to the interested reader of our thesis, we recommend the time investment of reading it from cover to cover. For we can in no way cover in a few paragraphs what Ashby took 295 pages to cover, and we will simply refer to many of the assertions from his text. 

Cybernetics utilizes elementary mathematical ideas to develop a basic framework for discussing feedback, stability, equilibrium, disturbance, regulation, information, entropy, noise, transmission (communication), constraints, and amplification to name a few. All these abstract tenets, when put in the context of the desired physical embodiment, can be applied to virtually any discipline of science, physics, engineering, biology, physiology, neurology, psychology, etc. In fact, the more complex the system, the more the derived tenets of cybernetics such as ``the Law of Requisite Variety'', help reduce the complexity of the `variety in disturbances' to a level that the system can be regulated even in the face of fixed regulation capacity \cite[pp. 246-247]{ashby1955introduction}. Hence, with our assertions yet to come in section \ref{sec:cybernetic_applications}, these principles apply to cyber security.

To close this section it seems important to lean toward how these abstract cybernetic tenets will be applied to the specific subject of cyber security. We believe that control theory, game theory, system theory, and information theory, when viewed through the lens of cybernetics, provide the tools to focus on breaking the cyber cycle. The next section will develop, as far as can be allowed in this forum, the basics of how these disciplines apply. Suffice it to say there are many corollary developments, which derive from this thesis, but can be published and discussed only in approved forums.


\section{Cybernetics Overview}
	\label{sec:cybernetics_overview}
	\subsection{Open Loop}
\begin{figure}[]
  \centering
  \includegraphics[width=0.48\textwidth]{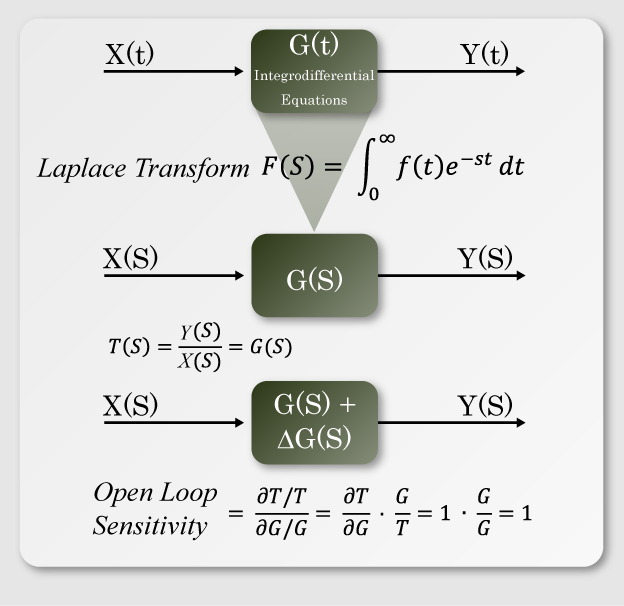}
  \caption{Open-loop System}
  \label{fig:open_loop}
\end{figure}

		Control theory is a branch of applied mathematics that focuses on dynamic systems. Systems have time-varying inputs $X(t)$ and outputs $Y(t)$, where the relationship between input $X(t)$ and output $Y(t)$ is governed by a transfer function of integral-differential equations $G(t)$ as seen in Figure \ref{fig:open_loop}. To simplify the math from solving calculus equations to solving algebraic equations, we generally evaluate these systems in the complex frequency domain, also known as the Laplace domain. Thus, the system becomes $Y(s)=X(s)G(s)$, where $G(s)$ represents the same transfer function. Such systems are defined as ``open loop'' systems because they do not use feedback to determine if the output has achieved a desired value. There are three cyber security related points to be made here about open-loop systems. First, in making the transformation to the Laplace domain, time has been removed from the analysis -- thus making the system stationary. Second, open-loop systems are highly sensitive to changes in the system's open loop transfer function $G(s) + \Delta G(s)$, as can be seen by the sensitivity equation in Figure \ref{fig:open_loop} and in texts on control theory \cite{nilsson2010eleccirc,rosenberg1983physys,dorf2011modern}. Third, it is easy to show that if the input is composed of an input signal and a disturbance or error signal: $X(s) = I(s) + D(S)$ then, $Y(s) = I(s)G(s) + D(s)G(s)$. Therefore, the transfer function has the same effect on the disturbance signal as the input signal.

\begin{figure}[]
  \centering
  \includegraphics[width=0.48\textwidth]{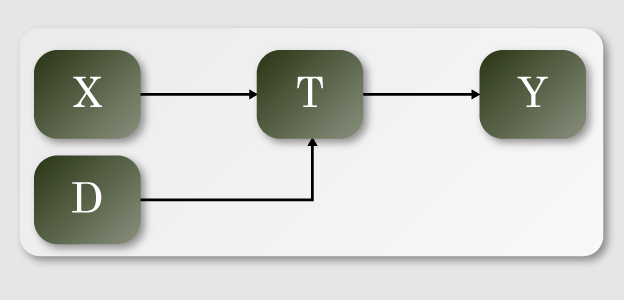}
  \caption{Cybernetic View of an Open-loop System}
  \label{fig:cybernetic_open_loop}
\end{figure}

Cybernetics, however, provides an alternative view of the open-loop system via a ``diagram of immediate effects'' \cite[pp. 57]{ashby1955introduction} The system and its susceptibility to parameter change and disturbance signals are shown in Figure \ref{fig:cybernetic_open_loop}. In this figure, there is a machine $X$ that provides an input to a ``transducer'' $T$ \cite[pp. 46]{ashby1955introduction}, which provides an output to a machine $Y$. Any parameter changes to the transducer are modeled as disturbance inputs and are lumped in with the other input disturbances, which were modeled as $D(S)$ above \cite[pp. 216-217]{ashby1955introduction}.

Despite the fact these two figures have some apparent diagrammatic similarities, they really do not convey the same meaning. The control theory viewpoint represents the transformation from one set of physical parameters to another as input to output (i.e., pressure to voltage, voltage to speed, speed to fluid density, voltage to temperature, etc.). The cybernetic viewpoint, however, is that information in the form of probability weighted ``Variety'' \cite[pp. 140]{ashby1955introduction}\cite[pp. 174]{ashby1955introduction} -- (also known as ``Entropy'') is being produced in machine $X$. This information is then combined with information from disturbance sources $D$ \cite[pp. 198]{ashby1955introduction} and then transformed from one representative form of information to another in the transducer $T$ \cite[pp. 140]{ashby1955introduction}. Finally, the information ends up as input at machine $Y$. The arrows of this diagram of immediate effects represent channels of communication that must exist for this system to operate \cite[pp. 210]{ashby1955introduction}. However, cybernetics fully expects these channels of communication are not always physical connections. Sometimes these connections are based on behavioral relations between endpoints \cite[pp. 180]{ashby1955introduction}.

Weiner defined \cite{wiener1948cybernetics} ``the amount of information'' or the entropy measured in bits as:

\begin{equation}
	H = \sum\limits_{i} {p_{i}\log_2(p_i)}
\end{equation}
Where $p_i$ is the probability of an event and,
\begin{equation}
	\sum\limits_{i} {p_{i} = 1}
\end{equation}

This is similar to work that Shannon did for information theory \cite{shannon1948comm} except for an insignificant minus sign that cancels out when calculating information gain. For an excellent explanation of why this minus sign is insignificant see \cite[pp. 177]{ashby1955introduction}. The importance of the idea of entropy here is that many of the properties of information theory apply to cybernetics. These will be leveraged further later.
	\subsection{Closed Loop}
\begin{figure}[]
  \centering
  \includegraphics[width=0.48\textwidth]{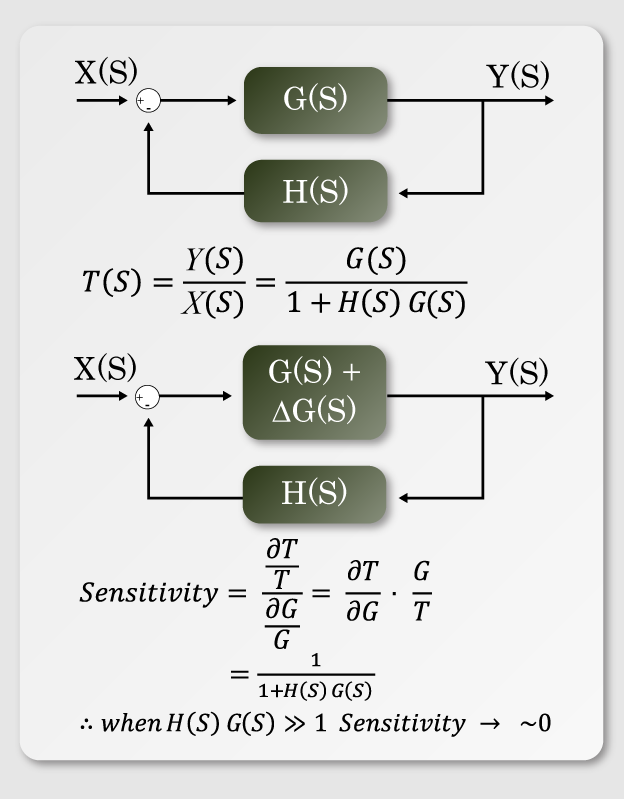}
  \caption{Closed-loop System}
  \label{fig:closed_loop}
\end{figure}

		As every undergraduate control theory student remembers, open-loop systems are quickly passed over for the more capable closed-loop or `error-controlled system', which is based on negative feedback. These systems look similar to those in Figure \ref{fig:closed_loop}, along with their stability equations. Some examples are the thermostat-controlled water bath, the air conditioning system, the cruise control, etc.

\begin{figure}[]
  \centering
  \includegraphics[width=0.48\textwidth]{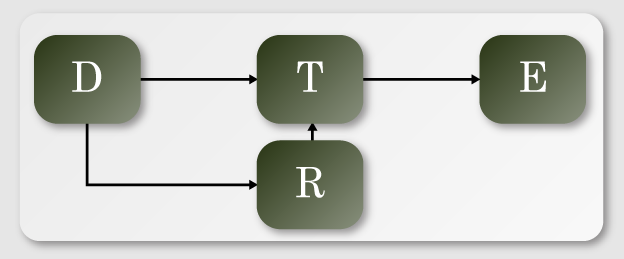}
  \caption{Cybernetic View of a Perfect Regulator}
  \label{fig:perfect_regulator}
\end{figure}

Given that cybernetics is the regulation of systems, one would expect Ashby's cybernetics text to offer an immediate discussion of closed-loop feedback systems. Surprisingly, however, this is not the case until near the end of his text. Instead, he focuses on something that is unrealizable and what he calls {\it the perfect regulator}, which he derives based on a game theory and system theory approach. This perfect regulator is unrealizable because it has the ability to counter the effects of the disturbance input in real-time as it is occurring. It is shown in the diagram of immediate effects in Figure \ref{fig:perfect_regulator}. The $D$ and the $T$ in this figure again represent disturbance sources and the transducer respectively as in Figure \ref{fig:cybernetic_open_loop}. The outcome $Y$ has been changed to $E$ to map more closely with Ashby's text and his examples, but no change in its meaning is intended. The $R$ represents this ``perfect regulator'' to which we have been referring. At first glance, it may not seem like an unrealizable system until one recognizes that in the case of the thermostat-controlled water bath, the regulator would have to react as described by Ashby as, ``I see someone coming with a cold flask that is to be immersed in me -- I must act now \cite[pp. 222]{ashby1955introduction}.''

So if the system is unrealizable, why does Ashby spend time discussing it at all? The answer is analogous to why our physics professors discussed the ``friction-less surface'', the ``mass-less spring'', or any of a host of other unrealistic scenarios; because so much can be learned from these simplified environments while still being applicable in more specific cases. 

Some additional cybernetics definitions and concepts we will use later are:
\begin{enumerate}
\item Variety is based on the permutations and/or combinations inherent in a set of distinguishable elements or inputs and can be measure in bits \cite[pp. 125-126]{ashby1955introduction}. 
\item Variety -- (and thus information) is not altered or lost by coding of a one-to-one transformation within transducers \cite[pp. 142]{ashby1955introduction} \cite[pp. 185-186]{ashby1955introduction}.
\item Information can be uniquely restorable to its original form after passing through several codes in succession provided that the variety in the set is preserved at every stage \cite[pp. 142]{ashby1955introduction}.
\item If a transform $U$ has an inverse $U^{-1}$, then both $U$ and $U^{-1}$ have one-to-one transformations where $U^{-1}$'s coding will restore the original message that was coded with $U$'s encoding \cite[pp. 142]{ashby1955introduction}.
\item The regulator $R$ -- (whether perfect or not) blocks and controls the flow of variety to perform its function \cite[pp. 199-201]{ashby1955introduction}.
\item Output variety in control cannot be less than  disturbance variety divided by regulation variety \cite[pp. 205]{ashby1955introduction}.
\item \#6 above said another way based on Logarithmic measure is $V_0 \ge V_D - V_R$, where $V_0$ is the variety of the outcome, $V_D$ is the variety of the disturbances and $V_R$ is the variety of the regulator. (This equation is known as the ``Law of Requisite Variety \cite[pp. 206-207]{ashby1955introduction}.'')
\item A corollary to \#6 and \#7 above is that only variety in $R$ can force down the variety due to $D$; in other words, variety can destroy variety \cite[pp. 206-207]{ashby1955introduction}.
\item Constraints -- (a condition that occurs when the variety that exists under one condition is less than the variety that exists under another) are the mechanisms used by regulators to reduce variety \cite[pp. 127]{ashby1955introduction} \cite[pp. 130]{ashby1955introduction}.
\item Every law of nature is a constraint \cite[pp. 130]{ashby1955introduction}.
\item $R$'s capacity as a regulator cannot exceed $R$'s capacity as a channel of communication \cite[pp. 211]{ashby1955introduction}.
\end{enumerate}

Given that some of these concepts about variety and constraints will be leveraged for our thesis shortly, it seems particularly important to give some illustrations to help bring the details to life. To do this, we will illustrate using a few examples from Ashby's text (modified slightly to improve clarity of context), which we will complete to help explain their meaning:
	\subsection{Examples}
		\label{sec:examples}
\noindent{\bf Example of Variety versus Constrained Variety \cite[pp. 133]{ashby1955introduction}}

Given a vector has ten components, where each component must be a member of the set $\{1,2,3,4\}$, answer the following:

\begin{enumerate}
\item How much variety does each component have?
\item How much variety exists for the possible set of vectors if the components vary independently?
\item How much variety exists if no two adjacent components may differ in value by more than one unit?
\end{enumerate}

{\noindent\emph Answers:}
\begin{enumerate} 
\item With the given set, each component can have 1 of 4 values so each has a variety of 4.
\item Our vector is of the form $< P_0, P_1, ..., P_9 >$, where $P_i$ are the members of the set $\{1,2,3,4\}$. Since order does matter, the variety of the set of possible vectors is equal to the permutation $4^{10} = 1048576$. This can also be written in bit notation: 
Variety $ = \log_2{4^{10}} = 20.0$ bits.
\item $21892$ or $\log_2{21892} = 14.4$ bits.
\end{enumerate}

{\noindent\emph Proof:}

\noindent For easier notation, take the equivalent problem of sequences of the letters $\{a, b, c, d\}$ subject to:
\begin{align*}
x_i &= a \rightarrow x_{i+1} = a \text{ or } b \\
x_i &= b \rightarrow x_{i+1} = a \text{ or } b \text{ or } c \\
x_i &= c \rightarrow x_{i+1} = b \text{ or } c \text{ or } d \\
x_i &= d \rightarrow x_{i+1} = c \text{ or } d
\end{align*}

\begin{align*}
\text{Let } A_i &\text{ be the number of such sequences of length } I \\
	&\text{ that start with the letter } a \\
\text{Let } B_i &\text{ be the number of such sequences of length } I \\
	&\text{ that start with the letter } b \\
\text{Let } C_i &\text{ be the number of such sequences of length } I \\
	&\text{ that start with the letter } c \\
\text{Let } D_i &\text{ be the number of such sequences of length } I \\
	&\text{ that start with the letter } d
\end{align*}

\noindent By the constraints, conclude that:
\begin{align*}
A_{i + 1} &= A_i + B_i \\
B_{i + 1} &= A_i + B_i + C_i \\
C_{i + 1} &= B_i + C_i + D_i \\
D_{i + 1} &= C_i + D_i
\end{align*}

\noindent With the base case of $A_1 = 1$, $B_1 = 1$, $C_1 = 1$, $D_1 = 1$: 

\noindent By induction, will show:
\begin{align*}
B_i = C_i = F_{2i} &\rightarrow \text{ Fibonacci of } 2i \\
A_i = D_i = F_{2i - 1} &\rightarrow \text{ Fibonacci of } 2i - 1
\end{align*}

\noindent This proves that the number of such sequences (total) is:
\begin{align*}
A_i + B_i + C_i + D_i &= 2(F_{2i} + F_{2i - 1}) \\
&= 2F_{2i + 1}
\end{align*}

\noindent Base case $i = 1$ is true because $F_1$ and $F_2$ are $1$, which is the same as $A_1$, $B_1$, $C_1$, $D_1$. 

\noindent Inductive case for A:
\begin{align*}
A_{i + 1} &= A_i + B_i = F_{2i - 1} + F_{2i} \\
&= F_{2i + 1} = F_{2(i + 1) - 1}
\end{align*}

\noindent Inductive case for B:
\begin{align*}
B_{i + 1} &= A_i + B_i + C_i = F_{2i - 1} + F_{2i} + F_{2i} \\
&= F_{2i + 1} + F_{2i} = F_{2i + 2} = F_{2(i + 1)}
\end{align*}

\noindent Inductive case for C:
\begin{align*}
C_{i + 1} &= B_i + C_i + D_i = F_{2i} + F_{2i} +  F_{2i - 1} \\
&=  F_{2i} + F_{2i + 1} = F_{2i + 2} = F_{2(i + 1)}
\end{align*}

\noindent Inductive case for D:
\begin{align*}
D_{i + 1} &= C_i + D_i = F_{2i} + F_{2i - 1} \\
&= F_{2i + 1} = F_{2(i + 1) - 1}
\end{align*}
\hspace{0.2in}Q.E.D

\noindent{\bf Example \#1 of Regulator's Channel Capacity \cite[pp. 211]{ashby1955introduction}}

A ship's telegraph from bridge to engineroom can determine
one of nine speeds not more often than one signal in five
seconds, and the wheel can determine one of fifty rudder
positions in each second. Since experience has shown that this
means of control is normally sufficient for full regulation,
estimate a normal upper limit for the disturbances (gusts,
traffic, shoals, etc.) that threaten the ship's safety.

\noindent Bridge to engine room control variety:
\begin{align*}
V_R(\text{Engine Room}) &= \frac{\log_2{(9\text{ speeds})}}{5\text{ secs}} \\
&= 0.63 \text{ bits/sec}
\end{align*}

\noindent Rudder regulation variety:
\begin{align*}
V_R(\text{Rudder Control}) &= \frac{\log_2{(50\text{ positions})}}{1\text { sec}} \\
&= 5.64 \text{bits/sec}
\end{align*}

\noindent Now from the Law of Requisite Variety ($V_0 \ge V_D - V_R$) and assuming the ship is in control:
\begin{align*}
0 \ge \sum{V_D} - \sum{V_R} \\ \sum{V_D} \le 6.3 \text{ bits/sec}
\end{align*}

\noindent{\bf Example \#2 of Regulator's Channel Capacity \cite[pp. 211]{ashby1955introduction}}

A general is opposed by an army of $10$ divisions, each of
which may maneuver with a variety of $10^6$ bits in each day.
His intelligence comes through $10$ signalers, each of whom
can transmit $60$ letters per minute for $8$ hours in each day, in a
code that transmits $2$ bits per letter. Is his intelligence channel
sufficient for him to be able to achieve complete regulation?

\noindent Individual division variety: 
\begin{align*}
V_D(\text{Division}) &= 10^6 \text{ bits/day} \\
\therefore V_D(\text{Army}) &= 10 * 10^6 = 10^7 \text{ bits/day}
\end{align*}

\noindent General's intelligence regulation channel:
\begin{align*}
V_R(\text{Channel}) = \text{ } &10 \text{ signalers } * 60 \text{ letters/min } * \\
					  &60 \text{ min/hour } * 8 \text{ hours/day } * \\
					  &2 \text{ bits/letter }\\
					= \text{ } &576000 \text { bits/day }
\end{align*}

The Law of Requisite Variety tells us that in order for the general's intelligence channel to be in control, $0 \ge V_D - V_R$. Because in this case $V_D \gg V_R$ , we know (to use Ashby's words) the General's intelligence channel ``is grossly insufficient.''

As we can see from the first example, the amount of variety was reduced by $5.6$ bits by simply constraining the environmental conditions. The next two examples of regulation showed systems where control was desired. The Law of Requisite Variety was used to show a bound for control in the former case based on the disturbance channel not exceeding $6.3$ bits/second. In the later case, the Law of Requisite Variety predicts the demise of the General's army because his channel for regulation is undersized by $17.36$ times. These examples, though simple in concept (and based on variety rather than entropy because of their equally probable inputs), lead exactly to the point of the cyber security problem, as we will discuss next.

		
\section{Cybernetic Applications to Cybersecurity}
	\label{sec:cybernetic_applications}

Now that we have laid an introductory foundation for cybernetics, we can focus on addressing the cyber cycle by using cybernetics for a new approach. 

	\subsection{Cyber Cycle}
\begin{figure}[]
  \centering
  \includegraphics[width=0.48\textwidth]{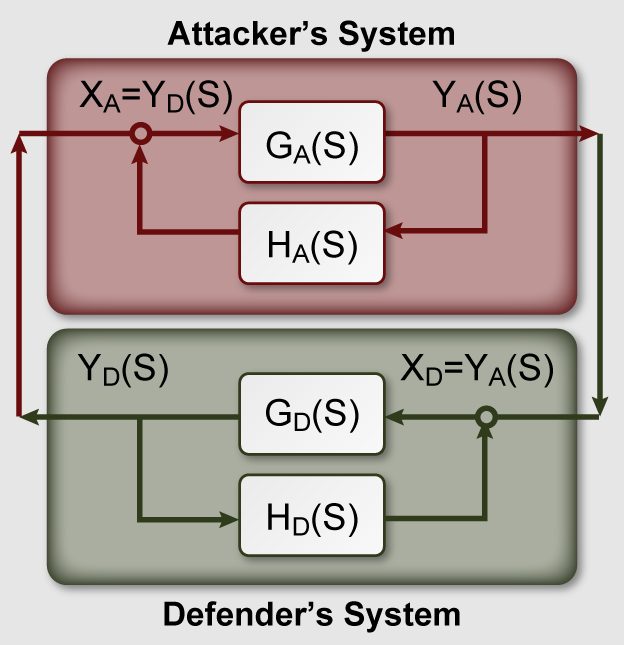}
  \caption{Cybernetic View}
  \label{fig:cybernetic_view}
\end{figure}

		As previously mentioned, new cyber security research must have a significant focus on breaking the continuous cyber cycle described earlier in Figure \ref{fig:cyber_cycle}. However, before we delve too deeply into this discussion, we first need to consider how the cyber cycle appears from a cybernetic perspective. A major difference is the attacker and defender's systems ($G_A(S)$ and $G_D(S)$ respectively) are closed-loop feedback systems with output that feeds into the input of the corresponding system as seen in Figure \ref{fig:cybernetic_view}. Notice, that we are not specifically defining what the system looks like from a theoretic systems standpoint; however, some examples could be a sensitive computer network, an embedded controller in a SCADA system or someone's personal electronics (laptop, smart phone and tablet). Also, the feedback loops and other communication channels within the systems may use behavioral relations rather than physical connectivity. Using this perspective, the main concept is that actions taken by either the attacker or defender (through each system’s regulator and transducer) will affect the input into the rival system. Examples of this include an attacker probing a system or a defender encrypting outgoing network traffic.

This cybernetic and control theory view of the cycle provides an excellent method to understand the cycle's continuous propagation. Initially, an attacker uses the defender's output to modify their output to negatively affect the defender. This could be any type of attack including Denial-of-Service or intrusion. Either way, it would cause some type of disturbance input to the defender's system, which would negatively affect that system and its output. Hopefully, the defender would detect the disturbances and attempt to mitigate those inputs through methods such as resetting their systems, adding network access control (firewalls) or developing patches. At this point, the system would return to its balanced state and the attacker would attempt to re-modify their system to create a new input disturbance. As long as the defender only responds to known and/or detected input disturbances, this cycle will continue indefinitely.

	\subsection{Stability}
		Again, since cybernetics was initially developed as a science of regulated systems, one of its primary concentrations is system stability. By viewing the attacker and defender through the lens of cybernetics, it would appear that one of the defender's primary goals would be to regulate an attacker's input to stabilize their defensive system's output. To better analyze this, consider a simplistic example of a public kiosk at a retail store used for online shopping. From a cybernetic standpoint, the system itself is a transducer that converts customer input to the final order. Examples of malicious or disturbance inputs could be an attacker modifying the system to steal customer information or making malicious orders from random accounts. Simplifying this, the system can have two states, clean or infected (variety of two), which would require a regulator capable of rebooting the system to a clean state whenever malicious activity is detected. The defender's goal would be to reset the system as soon as possible once malicious activity was detected to keep the system stable.

However, this simplistic example (while possible to implement) is far from an optimal real-world solution for a couple of reasons. First, the basic characterization of the kiosk's possible states is not sufficient to use as a model for current systems. For example, given enough time, a determined hacker could successfully compromise the system (even with locked-down permissions and read-only drives) and install malware that persists through reboots. At this point, the reboot process (regulator) becomes useless because the system will still be infected. From the cybernetic viewpoint, this would be characterized as variety the regulator could not account for because the system is too complex when taking into account behavioral channels. As systems become more complex and have more variety, the Law of Requisite Variety requires the corresponding regulator to have equal or greater variety in order to stabilize the system. In a calculus sense, this becomes an unbounded equation where: ($x$ refers to system complexity and the function refers to regulator complexity).

\begin{equation}
	 \lim_{x\to\infty}{F(x) = \infty}
\end{equation}

Continuing this example, the defender must be able to build a regulator that is capable of mitigating any attacks. From a real-world perspective, it is well known that as computers, networks, and software become increasingly complex, it is extremely difficult, if not impossible, to know every exploit, behavioral channel (social engineering) or attack vector that might exist. In reality, there is always some unknown vulnerability or new social engineering exploit that can introduce unknown variety into the system, meaning any regulator is deficient.

Secondly, the simplistic example assumes every attack can be detected (similar to above) and that it can be detected in enough time to reset the system without affecting legitimate users. As we just mentioned, some unknown exploit will always exist as systems become more complex, meaning any detector will be just as deficient as the regulator. Even if the system could detect the attack, to ensure stability it must be able to respond quickly enough to avoid affecting its users. In the case of the kiosk, a simple reboot may be sufficient on today's hardware, but most real-world systems require additional patching and other mitigation techniques. A good example is the recent hack and downtime of all of Apple's developer centers \cite{allthingsd2013apple}, which took three days to detect and over a week to mitigate.

Overall, this example is excellent proof of why current security techniques are ultimately lacking. From a cybernetic standpoint, defensive systems relying on a `detect then mitigate' strategy to build secure systems are inherently unstable. There will always be an attack that either remains undetected or causes significant downtime to patch, leading to instability. This example also demonstrates that, while it is a beneficial design model (further discussed in section \ref{sec:related_works}, a stability approach alone is not a feasible method to break the cyber cycle. Using stability, the cycle could be broken if a system was designed, such that it was capable of detecting any and all disturbance inputs (from an attacker) and counteracting them in real time. In cybernetics, this would be considered a perfect regulator and as Ashby alluded to in his work, this is unrealizable. This is consistent with the real-world fact that it is impossible to know every attack vector.
	\subsection{Constrained Variety}
		Even though stable, secure systems are difficult to build from a cybernetic viewpoint, one way to improve them and reduce their complexity is to use the concept of constrained variety. Essentially, the goal is to reduce the amount of disturbance variety (attacks) coming into the system. Based on the law of requisite variety, reducing disturbance inputs reduces the amount of control needed in order to regulate the system, which makes controlling the system more likely. In other words, the goal is to block attackers from exploiting the network and prevent users from accessing malicious content (even behaviorally). This role is currently fulfilled by systems such as firewalls, access control, VPNs and content filters. 

However, this constrained variety approach (again while  beneficial) is still vulnerable to the same unknown exploits and behavioral attacks as the stability approach. Reducing input disturbances could reduce the probability that a known attack or a probing type attack would be initially successful, but it does not eliminate it. The ultimate example of a system with constrained variety would be a sensitive computer network not physically connected to the Internet and only accessible by authorized personnel. However, as recent history has shown with Stuxnet \cite{wired2011stux} and Wiki-leaks \cite{wired2013blackbudget}, such systems are still vulnerable to behavioral channel exploits. In both cases, constraining the input variety to the system by isolating it from physical networks did not eliminate some behavioral channels and left the system vulnerable to eventual attack.
	\subsection{Instability and Non-stationarity}
		As we have seen, focusing strictly on constraining disturbance input and stabilizing the defender's system presents a very difficult security problem. However, if we expand our focus and take into account the actual communication channels between the attacker and defender (both inputs and outputs from the defender's perspective), we can see a different approach to breaking the cyber cycle. Previously, attempts to break the cycle focused on stabilizing the system and its outputs no matter what an attacker inputs. Defenders do not usually take into consideration the attacker's communication channel. However, if a system were to be designed, such that it and its communication channels (input and/or output as seen from the attacker's perspective) appear unstable, non-stationary or pseudo random, the attackers would face increasing difficulty trying to compromise the system. This type of system disrupts the cyber cycle by constantly adapting, thus making it difficult for attackers to even analyze, let alone communicate with the system.

If a defender's system $G(S)$ is encoded, then cybernetic theory predicts that this can be accomplished so that an inverse also exists. Using such an encoding for the system's input constrains the input variety in the cybernetic sense. This means that defenders can create specific channels that regulate or control the amount of variety allowed into the system. Additionally, if this is done in such a way that it is non-stationary, or pseudo chaotic, then $G(S)$ becomes $G(S,t)$. Thus, the system could be viewed as being encoded with a transform $U$ which eliminates malicious disturbance inputs because attackers lack the $U^{-1}$ secret decoder ring.

Many time-dependent attacks can be mitigated by encoding a system in a non-stationary way. For example, if a defender assumed that an attacker would eventually be able to find a behavioral channel of communication, such as insider threats, then they should also assume that an attacker could analyze their system or vulnerabilities at time $t < n$. Also, the defenders should be able to quantify a bound on the amount of entropy in their system by predicting probabilities of disturbance variety, based on historical incidents correlated with user access to the $U^{-1}$ transform. Using the predicted entropy and the Law of Requisite Variety, the defenders can modify their system so that once the attackers attempt to install malware, $\Delta G(s, t < n)$, the system will have moved to $G(S,t \ge n)$. Provided that the non-stationary change to $G(S,t \ge n)$ is sufficient to prevent installation and/or exploitation of the previously scanned vulnerabilities, the attackers will not be successful.

In addition, a non-stationary approach can be applied to both the input and output of the defender's system. Both approaches increase the complexity needed for an attacker's successful compromise of the defender's system. If we consider non-stationary output, such as rotating encryption keys, the attacker must be able to determine what keys are used and when they change. If their only option were to brute force attack each key and those keys are cycled before they can crack a single one, they would have a difficult time eavesdropping on communication. Another example of non-stationary input to a system would be a rotating authentication scheme similar to two-factor authentication. By constantly changing how a legitimate user finds and/or logs into the system, the attackers would have increasing difficulty even trying to discover or brute force attack the authentication system.

Cybernetics show that by increasing the possible variety in the defender's system and/or constraining input variety, the attackers must also increase the amount of variety in their system in order to regulate a communication channel between the two systems. For a constrained system, the attacker must increase his variety, either hoping to determine the inverse of encoding to access the system or to generate a new communication channel that bypasses those constraints. This is one reason phishing attacks are used so often; attackers are attempting to simply bypass the security without spending significant time analyzing the system. Using a non-stationary system, the defender forces attackers to handle extra variety (possible configurations) and disturbance inputs (parameter changes) while attempting to stabilize the system. However, today many systems are completely stationary, making an attacker's task far less complicated. The notion of stationary system development, though trivial to visualize in the complex frequency domain, is still easy to understand in the time domain. In essence, once built, the system components and/or parameters do not change. Once an attacker scans the system, they can develop malware and attack it (which can take considerable time depending on the system), confident the system has not been changed since the initial scan. Designed and implemented correctly, non-stationary systems eliminate this assumption for the attackers.
	\subsection{Breaking the Cyber Cycle}
		As we shift focus back to the goal of breaking the cyber cycle, it becomes apparent that many of today’s security solutions, even those with some amount of outer layers of non-stationarity, are simply stationary open-loop systems at their core. Many systems do not take into account the redeeming features of closed-loop systems and as a result are highly susceptible to disturbance inputs. For example, how often is a computer network topology designed as a feedback loop or how many CPUs use feedback to ensure memory interactions work as expected? In order to break the cyber cycle, security research needs to make a fundamental shift toward using cybernetic principles and the concepts of non-stationarity, stability and constrained-variety as the building blocks for secure systems.

In general, a cybernetic-influenced solution should incorporate all of these solutions, however non-stationarity will probably attract the most attention. The main purpose for building non-stationary systems is to dramatically increase the complexity for any attackers. This means that a system should be designed so that every component is non-stationary and constantly adapting to complicate tasks for the attacker. As more components of the system become non-stationary, the overall complexity increases according to the Law of Requisite Variety. The example of variety versus constrained variety in section \ref{sec:cybernetics_overview} is another way to consider this. Here, a single component of the vector has a variety of four. However, as components are combined to produce a large unique vector, the variety of the system grows exponentially. Our goal would be to design systems that follow the same principle; as more components are combined to uniquely define the system, the complexity grows exponentially \cite{cybersecrecommendations}. However, since the system must be usable, some components may be required to remain stationary, which is understandable. An excellent example would be a publically available webserver; without some stationary component, legitimate users would not be able to access it. Still, the goal for researchers is to implement non-stationary systems that are constantly behaving in an unpredictable manner to the attackers, but a predictable one to the users.

\begin{figure}[]
  \centering
  \includegraphics[width=0.3\textwidth]{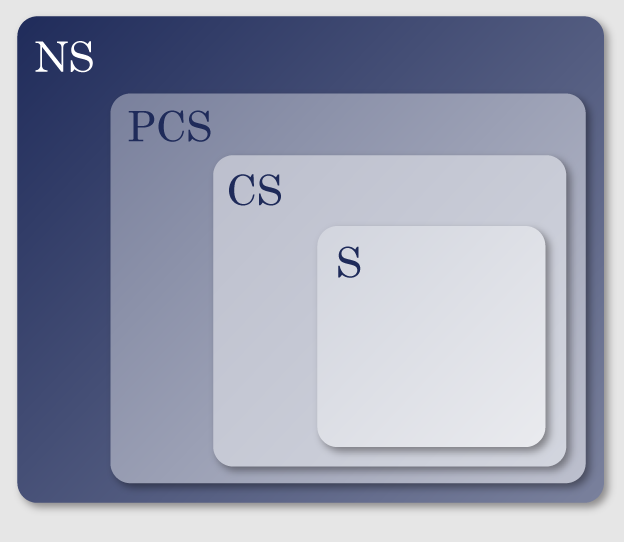}
  \caption{Relations of Stochastic Processes: Non-stationary, Poly-cyclostationary, Cyclostationary, and Stationary}
  \label{fig:stochastic_processes}
\end{figure}

In some cases, it could be difficult to generate a completely non-stationary process for a system. However, just as adding multiple layers increases complexity, interleaving multiple stationary processes together can result in the appearance of a non-stationary system. This is also known as a cyclo-stationary process, which is often used in communications. As a whole process appears non-stationary, it can easily be isolated into each subcomponent with a simple understanding of the cyclic nature of the interleaving. Figure \ref{fig:stochastic_processes} is a Venn diagram showing how a non-stationary signal (NS) can be contained within it: \cite{gardner1994cyclostationarity}

\begin{itemize}
\item A stochastic process (S) where $F_{X(t)} (x)$ is independent of the time translation parameter $t$,
\item A cyclostationary (CS) process with period T where $F_{X(t)} (x)$ is periodic in $t$ with period $T$,
\item A poly-cyclostationary (PCS) signal with period ${T} = T1,T2,T3,…$ Where $F_{X(t)} (x)$ is poly-periodic in $t$ with period ${T}$.
\end{itemize}

At this point, there will be those who recognize that there exist many solutions in academia and industry for creating ``moving target'' and diversity defenses that are very similar to our concept of non-stationary systems \cite{jajodia2011moving,jajodia2012moving}. In fact, the National Science and Technology Council (NSTC) and the Network and Information Technology Research and Development program (NITRD) have identified moving target defenses as a major focus for developing game-changing cyber security solutions \cite{cybersecrecommendations,cyberspace2011strategic}. These solutions can include randomization for memory addresses, instruction sets, encryption keys, network configurations, virtual machines, operating systems, etc. However, while many of these solutions employ some of the properties discussed above, not all are sufficient to break the cyber cycle and be classified as non-stationary. In fact, when considering methods for breaking the cyber cycle, moving target solutions should be considered as a tool for implementing non-stationary systems rather than being a solution on their own. The following paragraphs help provide clarity why:

\begin{enumerate}
\item Most moving target defense research is focused on building a specific technology or implementation rather than considering an entire system. Narrowing the focus too much to specific technology ignores the fact that behavioral channels or other attack vectors can completely bypass the defense, rendering it useless (further discussed in section \ref{sec:related_works}) \cite{jajodia2011moving}.

\item Many systems have not deployed non-stationary solutions throughout, but instead it is argued that implementing moving targets outside of a system, as a hardened shell, is a sufficient solution. Simply adding a moving target to the outside of a stationary system does not ensure the system is secure. In addition, single layer moving target defenses do not provide adequate protection. For example, in 2011, a well-known and respected randomization token provider for two-factor authentication was hacked. This hack cost the company millions to issue new tokens. It was not the randomization token that was hacked, but rather the stationary aspects of their internal solution. In these cases, the system behind the moving target defense was stationary leaving the system vulnerable to behavioral channels. 

The application of cybernetics to this problem suggests that unless the entire solution is encoded to constrain incoming communication channels (regulated), then a disturbance variety source will find a path that bypasses the shell completely. Once a behavioral channel was exploited, the hackers were able to compromise the system itself without regard to penetrating the two-factor authentication. Therefore, the moving target solution becomes irrelevant. Therefore, given enough time such a path can be exploited, even if it has an extremely low probability of being found.

\item Most moving target solutions offer no mathematical foundation for predicting when a system should change. Some use a set frequency (such as 24 hours); while others simply state the system must move twice the time it takes an attacker to compromise a stationary system \cite{beraud2011using} based in some rudimentary form on the Nyquist sampling theorem. These values are completely subjective and depend on the type of attack and the target. However, cybernetics and Law of Requisite Variety \cite{ashby1955introduction} predict the theoretical regulation requirement for preventing cyber intrusion based on variety and/or entropy. This is analogous to how entropy predicted Shannon's channel capacity and coding theory \cite[pp. 183-186]{ashby1955introduction}.

\begin{equation}0\ge\sum{H_D}  - \sum{H_R}\end{equation}

Here, $H_D$ is the disturbance entropies (probabilistic varieties) allowed to enter the system of interest, and $H_R$ is the regulation system's entropies employed to constrain the initial disturbances. From this equation, it is easy to see that systems fully exposed to high bandwidth channels (i.e. the Internet) pose great cyber security risk unless constrained. This is because the amount of disturbance variety has no upper bound. Additionally, regulation entropies are often fixed, especially with existing systems, since non-stationary system development is not a widespread trend today.

\item Some solutions, such as uniquely compiled applications and address-space layout randomization, do not continuously adapt their configuration during runtime \cite{jajodia2011moving,grsecurity}. In the case of compiler-generated diversity, the randomization occurs as a user downloads applications from a store; address randomization only occurs during task creation, meaning long running tasks are not constantly randomized. These type of defenses are most effective in blocking wide spread attacks against static addressing and binaries. However, over the long run, it still leaves individual systems vulnerable to attack. Given enough time, determined attackers can discover ways to exploit vulnerabilities in these stationary systems despite the trivially implemented layers of ``moving target defense''.  In other words, a moving target system that is implemented, but never changed, might as well be a stationary system given enough time.

\item Though it is outside the scope of this paper, one final reason is that {\it moving target defense} research is specifically focused on defensive implementations. On the other hand, non-stationary systems can involve far more than just defensive applications.

\end{enumerate}

While moving target defenses alone are not sufficient to break the cyber cycle, when combined with the principles of cybernetics, we can get a better picture on how cyber systems should be designed. One of the major benefits that cybernetics offers to the less-than-optimal solution of using layered moving target defense is predicting an upper bound on the amount of time needed before a change is recommended. This would be based on the variety of all known interactions with the system (both human and machine) plus a sufficient safety margin for unknown channels.

	\subsection{Addressing Reliability and Usability}
		When considering how to build non-stationary systems that break the cyber cycle, it is paramount to ensure the current requirements of normal operations avoid negative impact. The ability to maintain normal operations requires addressing the system's ability to provide a reliable and predictable system with minimal impact to usability.

An entire architecture that is completely non-stationary would likely require implementing Ashby's perfect regulator. Therefore, to provide enough predictability to allow normal behavior, a hybrid approach is assumed to be necessary at some level. That being said, however, enough layers of entropy must be present and leveraged to diminish an attacker's attempts. Additionally, reducing stationarity at the system's core will cause more difficulty for the attackers and reduce the chances a bypass behavioral channel will be successful. This is the balance that follow-on applied research of these cybernetic ideas must achieve. We recommend future research examine areas such as:

\begin{enumerate}
\item The theoretical limits to which non-stationary functions can be utilized in a system.
\item The time variance and entropy that can be introduced by a non-stationary system in order to maintain reliable communications.
\item Modeling and simulation of typical and common systems to verify theoretical analysis.
\item The impacts of cybernetics on closed-loop systems and closed-loop system properties such as: instability, time delay, root locus, etc.
\item The impacts of cybernetics on advanced control theory topics such as Kalman filters.
\item The creation of new forward and inverse non-stationary transforms that provide stability for normal operation.
\end{enumerate}


\section{Related Applications}
	\label{sec:related_works}

As discussed in the previous section, there are many examples of ongoing applied research based on an intuitive perspective of cybernetics. The purpose of this section is to provide a limited set of specific examples of the broader assertions made previously, and to show how cybernetics predicts the limitations of these approaches. Additionally, we will discuss one system whose technology development was based in the ideas of non-stationarity. Therefore, it offers a portion of the variety needed to create a highly non-stationary network solution without making the claim that it is a complete solution for breaking the cyber cycle.

	\subsection{Effectiveness of Moving Target Defenses}
		In the second chapter of ``Moving Target Defense: Creating Asymmetric Uncertainty for Cyber Threats'', Evans et al. address the effectiveness of common diversity defenses against several types of attacks \cite{jajodia2011moving}. They specifically focus on implementations of address space randomization, instruction set randomization and data randomization. These defenses are designed to complicate attacks by randomizing data locations or program instructions. In addition, they develop a probability model for successful attacks, which is similar to ours, based on entropy in the defender's system.

Using this model, they evaluate the effectiveness of each defense against several attack types: circumvention attacks, deputy attacks, brute force, entropy reduction, probing attacks and incremental attacks. Based on their analysis, layout randomization defenses have no advantage defending against circumvention and deputy attacks, since these attacks are specifically designed to bypass the randomization . In the case of deputy attacks, the attacker uses other legitimate software on the system to bypass the randomization. In the case of entropy reduction, these defenses may at most double the attack duration before it is successful. For the last two attacks, probing and incremental, diversity defenses may provide a distinct advantage if the randomization is rapid enough.

Based on their evaluations, a good observation to make is that diversity defenses must continuously be randomized in order to be effective against attacks. In the case of address space layout randomization, many current implementations provide limited randomization to the base address for the executable. Also, this randomization usually only occurs during process creation \cite{grsecurity}. Because of this, an attacker can circumvent these defenses for long running tasks since they do not change over time.

	\subsection{Resilient Web Services}
		Another good moving target example to analyze is an implementation for supporting web services \cite{huang2010security,huang2009automating,jajodia2011moving}. Overall, the system was designed using a control theory approach intended to keep a pool of virtual servers in their stable state. The system used a closed-loop control circuit between the pool of virtual servers, network sensors and a controller for detecting attacks and resetting affected servers to their initial state in order to wipe out any malicious activity on the system. In addition, diversity is added to the system by randomizing the configurations of installed software for each virtual machine. A dispatcher takes requests from the public Internet and routes it to a clean virtual machine to serve the request.

This design presents a good example of stability from a cybernetic perspective. Resetting affected servers to an initial install state allows the system to be kept in a relatively stable state to break up an attacker's communication channels. However, even though this system is a moving target and does have some non-stationary properties, it is not completely equivalent to a non-stationary system. While it does employ a level of diversity with the web-server software stack, the number of possible configurations is relatively low. If an attacker finds a specific configuration and develops an attack offline, it would only be a matter of time before that configuration serves a request (especially with a large botnet sending several attacks). Once successful, the attacker does need to contend with server resets. This does provide a level of non-stationarity for the attacker (since it interrupts their control), but this is more a stability aspect of the system. Given enough time, an attacker could develop methods to attack any configuration and attempt to compromise the underlying virtualization rather than the servers themselves. In addition, the system is not moving its address on the Internet, leaving it vulnerable to Denial of Service attacks. A truly non-stationary system would constantly be moving its logical location making it harder for an attacker to find, much less attack.
	\subsection{Hypervisor for Web Browsing}
		Legacy systems connected to a network are considered stationary and can be infected by a variety of viral software constructs, generally referred to as malicious software or malware. Without the knowledge of the user, malware can be unintentionally downloaded to a computer allowing for the installation of various types of subservient software. Many of these system exploits are undetectable by commercial antivirus solutions and will survive a system reboot, thus providing an attacker with persistent control over the computer while the system owner is completely oblivious to the compromised system.

L-3 Communications National Security Solutions (STRATIS) has developed a sophisticated, yet simplistic, virtualized hardened solution to break the attack chain on stationary systems. The solution creates a hardened virtual barrier between individual, ``guest'' nodes accessing ``at risk'' services such as the Internet and their host network. The architecture has been developed to block the ability for a hosted system on a trusted local area network (LAN) to directly communicate to an untrusted environment (e.g., Internet). Access to untrusted environments is provided by the virtual node, without the risk of compromise to the host or corresponding trusted host network.

The virtual machine can access an external system but cannot communicate directly to resources on the trusted network (e.g., LAN/WAN). The solution provides a scalable system that can be applied to a variety of system architectures or hardware configurations, supporting traditional bare metal and virtual desktop infrastructure systems.

The host computer operates a hypervisor logically separating the virtual machine from the host computer's operating system. The secure virtual abstraction layer operates a modified application (e.g., Internet browser) that connects the system through an encapsulated and restricted communication path to the untrusted environment. The virtual guest operates internally as a normal system, except that it relies on the host connection as a transport. The guest appears as a single node machine such as a public kiosk or home PC and contains no intellectual property or data of value to an adversary.

From a cybernetic perspective, this solution provides both non-stationarity and stability for the system. Stationarity is interrupted and resiliency is increased by utilizing a security-hardened, revertible system that provides enhanced anomaly detection that takes action when dangerous data has been encountered by the guest operating system. If compromised, the guest environment is automatically restored from a known, clean image, removing all infections and disrupting any remote control and exfiltration channels that break any persistence by the adversary. In addition, it provides stability for the user by restoring the virtual system when anomalies are detected or at the users' time discretion. Overall, this results in a non-stationary system with a feedback loop to the host system.


\section{Conclusion}
	\label{sec:conclusion}
	The goal of cyber-security should be to develop multilayered, non-stationary systems that meet certain criteria. From the instability perspective, they should be unpredictable and highly complex -- forcing attackers to have complex systems. They must be constantly adapting to both: 1) break the attacker's ability to follow the cyber cycle; 2) disrupt any behavioral channels that outer layers may not be capable of limiting. From a stability perspective, they must be able to keep the system internally clean and provide constrained input into the system. While the outside of the system produces complex, non-deterministic disturbance inputs into an attacker's control system, the interior of the system must provide constrained inputs and stable, regulated operation. We see this as fertile ground for many follow-on research topics and applications.

\section*{Acknowledgment}

The authors would like to thank the many technical reviewers at L-3 for their diligence in ensuring accuracy including Douglas Bowen and Randal Sylvester. Many thanks to Martin T. Jakub from L-3 CSW for the proof in section \ref{sec:examples}. We would also like to recognize Sonya Rowe for all her assistance in managing the non-technical requirements that kept the project moving along efficiently so that we could concentrate on developing the new concepts and writing.



\bibliographystyle{IEEEtran}
\bibliography{References}

\end{document}